\documentclass{ws-ijmpb}

\usepackage{url}
\usepackage{graphicx}
\usepackage{dcolumn}
\usepackage{bm}
\usepackage{hyperref}
\usepackage{xcolor}

\begin{document}

\markboth{A. Sourie \& N. Chamel}
{Hydrodynamic forces acting on a quantized vortex}

%
\catchline{}{}{}{}{}
%

\title{GENERALIZATION OF THE KUTTA-JOUKOWSKI THEOREM FOR THE HYDRODYNAMIC FORCES ACTING ON A QUANTIZED VORTEX}

\author{AUR\'ELIEN SOURIE\footnote{Also at LUTH, Observatoire de Paris, PSL Research University, CNRS, Universit\'e Paris Diderot,  Sorbonne Paris Cit\'e, 5 place Jules Janssen, 92195 Meudon, France.}}
\address{Universit\'e Libre de Bruxelles, Physics Department, CP-226, 
B-1050 Brussels, Belgium\\
asourie@ulb.ac.be}

\author{NICOLAS CHAMEL}

\address{Universit\'e Libre de Bruxelles, Physics Department, CP-226, 
B-1050 Brussels, Belgium\\
nchamel@ulb.ac.be}

\maketitle

\begin{history}
\received{Day Month Year}
\revised{Day Month Year}
\end{history}

\begin{abstract}
The hydrodynamic forces acting on a quantized vortex in a superfluid have long been a highly controversial issue. A new approach, originally developed in the astrophysical context of compact stars, is presented to determine these forces by considering small perturbations of the asymptotically uniform flows in the region far from the vortex in the framework of Landau-Khalatnikov two-fluid model. Focusing on the irrotational part of the flows in the Helmholtz decomposition, the classical Kutta-Joukowski theorem from ordinary hydrodynamics is thus generalized to superfluid systems. The same method is applied to predict the hydrodynamic forces acting on vortices in cold atomic condensates and superfluid mixtures. 
\end{abstract}


\section{Introduction}
\label{sec:intro}

One of the most remarkable manifestations of superfluidity and superconductivity is the formation of topological defects. First directly observed\cite{essmann1967,yarmchuk1979} in  a lead-indium alloy, niobium and helium-4, individual quantized vortices have been routinely detected in many other superfluid and superconducting systems. Conversely, their presence in recently produced dilute atomic condensates has been interpreted as a proof of superfluidity (see, e.g., Refs.~\refcite{abo-shaeer2001,zwierlein2005,yao2016observation}). 

The dynamics of vortices plays a crucial role in various phenomena, such as the dissipation of electric currents in type-II superconductors or the development of quantum turbulence in superfluids.\cite{donnelly2005,mangin2017} However, a complete understanding of the forces governing the motion of vortices is still lacking. The situation remains particularly unsatisfactory for the emblematic case of superfluid helium at finite temperatures. According to the two-fluid model\cite{khalatnikov1989introduction}, the superfluid with mass density $\rho_{\,\text{s}}$ and velocity $\pmb{v_{\,\text{s}}}$ coexists with a ``normal'' viscous fluid  with mass density $\rho_{\,\text{n}}$ and velocity $\pmb{v_{\,\text{n}}}$. The existence of a relative superfluid flow round a vortex gives rise to a force (per unit length) of the Magnus type\cite{wexler1997}
\begin{equation}\label{superfluid-magnus}
\pmb{\mathcal{F}_{\text{sM} }} = - \bar{\rho}_{\, \text{s} }\, \kappa \, \pmb{\hat{z}} \times \pmb{\bar{v}_{\, \text{s} }}\, , 
\end{equation}
where $\bar{\rho}_{\, \text{s} }$ and $\pmb{\bar{v}_{\, \text{s} }}$ denote the asymptotically uniform  superfluid density and velocity respectively, $\kappa=h/m$ is the quantum of circulation ($h$ being the Planck constant and $m$ the mass of the superfluid particles), and $\pmb{\hat{z}}$  is a unit vector in the direction of the vortex. This force has been experimentally confirmed by measurements of a vortex trapped on a wire.\cite{vinen1961,whitmore1968,zieve1993} The additional forces induced by the normal-fluid flow have been highly controversial (see, e.g., Refs.~\refcite{putterman74superfluid,thouless1996tranverse,kopnin2002vortex,sonin2010} and references therein). These forces have been most often empirically parametrized as
\begin{equation}\label{drag-force}
\pmb{\mathcal{F}_{\text{n} }}= -\gamma_0\, \pmb{\hat{z}}\times (\pmb{\hat{z}}\times \, \pmb{\bar{v}_{\, \text{n} }})+\gamma_0^\prime\,  \pmb{\hat{z}}\times \pmb{\bar{v}_{\, \text{n} }}\, ,
\end{equation}
where $\pmb{\bar{v}_{\, \text{n} }}$ denotes the asymptotically uniform velocity of the normal fluid. 
The normal forces have been alternatively described in terms of the relative velocity $\pmb{\bar{v}_{\, \text{n} }}-\pmb{\bar{v}_{\, \text{s} }}$ as (see, e.g., Ref.~\refcite{donnelly2005})
\begin{equation}\label{drag-force2}
\pmb{\mathcal{F}_{\text{n} }}= -\alpha\rho_{\, \text{s} }\kappa\, \pmb{\hat{z}}\times (\pmb{\hat{z}}\times \, (\pmb{\bar{v}_{\, \text{n} }}-\pmb{\bar{v}_{\, \text{s} }}))-\alpha^\prime\rho_{\, \text{s} }\kappa\,  \pmb{\hat{z}}\times (\pmb{\bar{v}_{\, \text{n} }}-\pmb{\bar{v}_{\, \text{s} }})\, .
\end{equation}
The coefficients $\gamma_0$, $\gamma_0^\prime$ and $\alpha$, $\alpha^\prime$ have been experimentally found to have some dependence on the temperature as well as on the relative velocities.\cite{donnelly2005,barenghi1983friction,bevan1995vortex,bevan1997vortex} 
Second-sound experiments seem to suggest the existence of a small axial friction force directed along the vortex line.\cite{mathieu1984} Nevertheless, this force, whose origin remains poorly understood, is usually neglected (see, e.g., Section 3.6 in  Ref.~\refcite{donnelly2005}). Different microscopic interpretations of the forces~\eqref{drag-force} or \eqref{drag-force2} have been proposed.\cite{kopnin2002vortex,sonin2016} However, comparisons with experiments have been hampered by the matching to macroscopic scales. In any case, the microscopic theory should reproduce the superfluid hydrodynamic equations, since the latter are based solely on fundamental conservation laws, thermodynamic principles, and symmetries, as pointed out in Ref.~\refcite{sonin2016} (see also Ref.~\refcite{thouless1996tranverse}). The alternative approach is therefore to rely on the two-fluid model. The two-fluid hydrodynamic equations are generally linearized and solved in the vicinity of the vortex, where the normal-fluid flow is governed by viscous effects.\cite{mathieu1980hydro,thouless2001vortex,sonin2001LNP} The flows are usually further assumed to be incompressible. However, such an analysis faces two serious problems: (i) the hydrodynamic description breaks down at short length scales (typically of the order of the mean-free path of the excitations), (ii) the inner boundary conditions to impose on the hydrodynamic equations remain uncertain since the vortex is not an external body immersed in the superfluid but is part of the flow,  as stressed, e.g., in Ref.~\refcite{putterman74superfluid}. The resulting expressions for the forces thus depend on somewhat arbitrary cutoffs. As a matter of fact, the analyses of Refs.~\refcite{mathieu1980hydro,thouless2001vortex,sonin2001LNP} lead to different predictions. 

To avoid the pitfalls that plague previous studies, we propose in this paper a new treatment, originally developed in the  astrophysical context of compact stars, consisting in solving the hydrodynamic equations only in the region far from the vortex, where the perturbations of the asymptotically uniform flows are small. To illustrate the method, we have considered in Section~\ref{sec:forces} the two-fluid model of superfluid helium developed by Landau and Khalatnikov.\cite{khalatnikov1989introduction} Making minimal and standard assumptions on the flows, we have thus generalized the classical Kutta-Joukowski formula\cite{kutta1902,jouko1906} to superfluid systems. Results are discussed and summarized in Section~\ref{sec:conclusions}, where it is shown that the same approach can be followed to determine the forces acting on vortices in cold atomic condensates and superfluid mixtures.


\section{Generalized Kutta-Joukowski theorem}
\label{sec:forces}

\subsection{Hydrodynamic forces on a vortex: general definition}
\label{sec:definitions}

\begin{figure}[bt]
	\centerline{\psfig{file=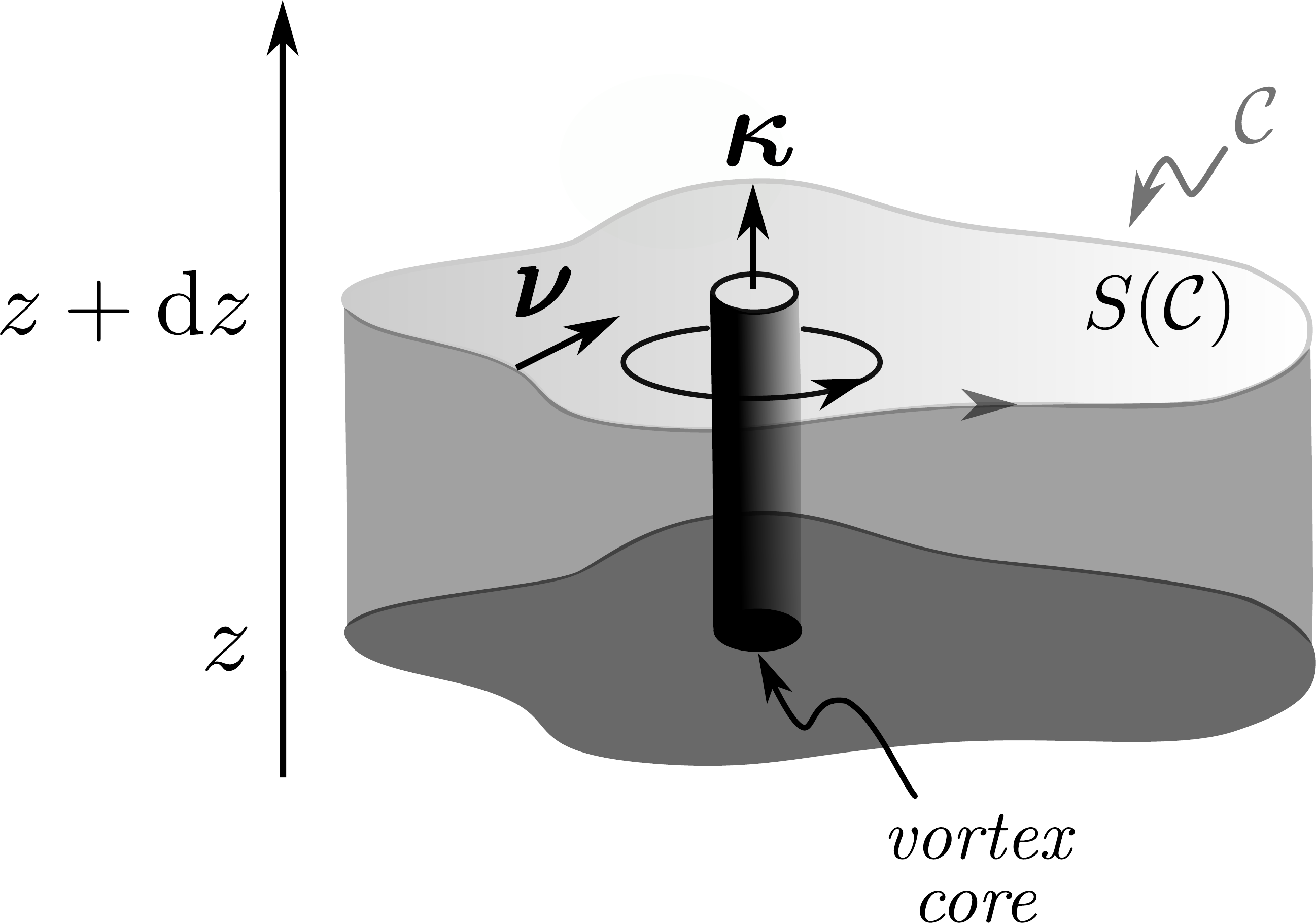,width=3in}}
	\vspace*{8pt}
	\caption{Schematic picture illustrating the fluid element contributing to the force per unit length acting on the vortex. See text for details.}
	\label{fig:force}
\end{figure}

Let us consider an asymptotically uniform vortex configuration which is both in steady-state and longitudinally invariant, say along the $z$ axis. The force density $\pmb{f}$ acting on a fluid particle is defined by the divergence of the momentum-flux tensor $\Pi^{ij}$ ($i$, $j$ denoting space coordinate indices, and adopting Einstein's convention that repeated indices are summed):  
\begin{equation}
f_i\equiv\nabla_j \Pi^j_i \, .
\end{equation}
The force $\pmb{\text{d}F}$ exerted on a vortex segment of length $\text{d}z$ by a fluid element whose  volume  is delimited by a closed contour $\mathcal{C}$ encircling the vortex, as represented on Fig.~\ref{fig:force}, is thus given by
\begin{align}\label{def-force}
	\text{d}F_i&=-\iiint f_i\,  \text{d}V=-\iiint \nabla_j \Pi^j_i \, \text{d}V \nonumber \\
	&=\iint_{S(\mathcal{C})} \Pi^j_i(z)\hat{z}_j\,\text{d}S-\iint_{S(\mathcal{C})} \Pi^j_i(z+\text{d}z)\hat{z}_j\, \text{d}S \nonumber \\
	&+\text{d}z \oint_\mathcal{C} \Pi^j_i\nu_j \, \text{d}\ell \, ,
\end{align}
making use of Stokes' theorem, and where $\pmb{\nu}$ is a unit vector perpendicular to both the vortex line and the contour $\mathcal{C}$, and is oriented inside the contour (see Fig.~\ref{fig:force}). Longitudinal invariance along the vortex line implies that $\Pi^{ij}$ is independent of $z$. The two surface integrals in the second line of Eq.~\eqref{def-force} thus cancel each other. The force per unit length acting on the vortex can be finally expressed as 
\begin{equation}\label{jouko-def}
\mathcal{F}_{i}\equiv\frac{\text{d}F_i}{\text{d}z}=\oint_\mathcal{C} \Pi^j_i\nu_j \, \text{d}\ell \, .
\end{equation}

\begin{figure}[bt]
	\centerline{\psfig{file=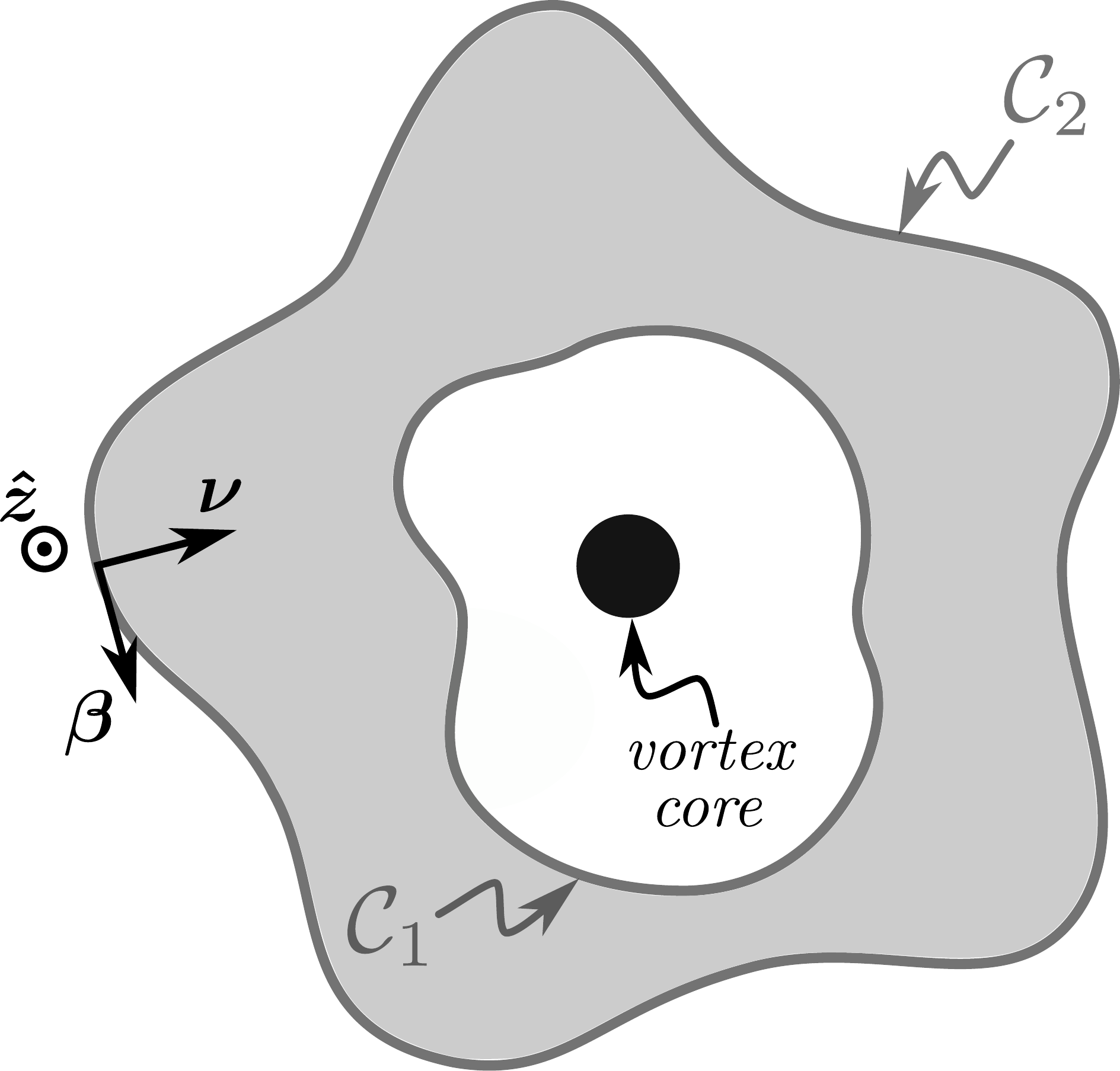,width=2.5in}}
	\vspace*{8pt}
	\caption{Schematic picture illustrating the surface $\mathcal{S}(\mathcal{C}_2)\, \backslash\, \mathcal{S}(\mathcal{C}_1)$ (shaded area) delimited by two different contours $\mathcal{C}_1$ and $\mathcal{C}_2$ around the vortex.}
	\label{fig:surface}
\end{figure}

The force \eqref{jouko-def} is well-defined provided the contour integral is evaluated at sufficiently large distances from the vortex where the force density vanishes, $f_i=0$. Indeed, considering two different contours $\mathcal{C}_1$ and $\mathcal{C}_2$, as illustrated in Fig.~\ref{fig:surface}, we have 
\begin{equation}
\mathcal{F}_i(\mathcal{C}_1)-\mathcal{F}_i(\mathcal{C}_2) = 
\iint_{\mathcal{S}(\mathcal{C}_2)\, \backslash\, \mathcal{S}(\mathcal{C}_1)} f_i\,  \text{d}S= 0\, ,
\end{equation}
therefore $\mathcal{F}_{i}(\mathcal{C}_1)=\mathcal{F}_{i}(\mathcal{C}_2)$.

\subsection{Hydrodynamic equations far from the vortex}
\label{sec:hydro-eqs}

Although the hydrodynamics of a single vortex has been already studied, e.g., in Refs.~\refcite{mathieu1980hydro,thouless2001vortex,sonin2001LNP}, we focus here on a region sufficiently \emph{far from the vortex} for the first-order perturbation theory to hold. Asymptotically, the fluids flow uniformly and their velocities are thus given by $\pmb{\bar{v}_{\, \text{n}}} $ and $\pmb{\bar{v}_{\, \text{s}}} $. We can always consider distances sufficiently far from the vortex for the velocities to be expressible as  $\pmb{v_{\, \text{n}}} = \pmb{\bar{v}_{\, \text{n}}} + \delta\pmb{v_{\, \text{n}}}$ and $\pmb{v_{\, \text{s}}} = \pmb{\bar{v}_{\, \text{s}}} + \delta\pmb{v_{\, \text{s}}}$, where $\delta \pmb{v_{\, \text{n}}}$ and $\delta\pmb{v_{\, \text{s}}}$ denote small disturbances of the uniform background flows. Similarly, the momentum-flux tensor can thus be expressed as 
\begin{equation}
	\Pi_{ij} =  \bar{\Pi}_{ij} + \delta \Pi_{ij}\, ,
\end{equation}
where $ \delta \Pi_{ij}$ denotes a small disturbance of the uniform background momentum-flux tensor $\bar{\Pi}_{ij} $. Since the force integral for the unperturbed uniform background flows must evidently vanish by symmetry, the corresponding value in the presence of the vortex~\eqref{jouko-def} will be given to first order by 
\begin{equation}\label{jouko-def2}
	\mathcal{F}_{i} = \oint_\mathcal{C} \nu_j \, \delta \Pi^j_{\, i} \, \text{d}\ell\, .
\end{equation}

According to Helmholtz's theorem, the perturbation of the normal velocity field $\delta\pmb{v_{\, \text{n}}}$ can be uniquely decomposed into an irrotational part and a solenoidal part as (see, e.g., Ref.~\refcite{aris1989})
\begin{equation}
\label{helmholtz}
\delta \pmb{v_{\, \text{n}}}=\pmb{\nabla} \Phi +\pmb{\nabla}\times \pmb{W} \, .
\end{equation}
The superflow (far from the vortex) is purely irrotational by nature, 
\begin{equation}
\pmb{\nabla}\times\delta\pmb{v_{\, \text{s}}}=\pmb{0}\, .
\label{irrot_sup}
\end{equation}
Because the hydrodynamic equations far from the vortex are linear in the perturbed velocities, the general solution can be obtained by considering separately the two contributions to the normal flow in Eq.~\eqref{helmholtz}. In what follows, we will focus on the solution corresponding to the first term in the Helmholtz decomposition~\eqref{helmholtz}. This means that we will consider here that the perturbed normal velocity can be written as $\delta \pmb{v_{\, \text{n}}}= \pmb{\nabla} \Phi $ such that 
\begin{equation}
\pmb{\nabla}\times\delta \pmb{v_{\, \text{n}}}=\pmb{0}\, ,
\label{irrot_normal_vel}
\end{equation}
as in the classical derivation of the Kutta-Joukowski theorem. We leave the calculation of the full solution for future studies. 

As in previous studies on superfluidity (see, e.g., Ref.~\refcite{thouless2001vortex}) as well as in the classical derivation of the Kutta-Joukoswki theorem, we shall neglect the spatial variations of the mass density $\rho=\rho_{\, \text{n}}+\rho_{\, \text{s}}$ and of the entropy density $s$. The former approximation is very well justified for superfluid helium-4 given the very small values for the thermal expansion and isothermal compressibility coefficients, respectively of order $\alpha\sim 10^{-4}-10^{-2}$~K$^{-1}$ and $k_T\sim 10^{-8}$ cm$^{2}$~dyn$^{-1}$ (see, e.g., Refs.~\refcite{elwell1967,harrislowe68}). The latter approximation is more restrictive. Although the entropy density $s$ is very weakly dependent on the pressure, it varies appreciably over the whole temperature range for which helium-4 is superfluid (see, e.g., Ref.~\refcite{meijdenberg1961}). But contrary to previous studies, the assumption of uniform entropy density is only made here in the region far from the vortex where temperature fluctuations remain small. Considering that the relative velocity of the normal and superfluid motions is typically small compared to both the speed of sound $c$ and the ratio $k_\text{B} T / p_0$ (where $k_\text{B}$ is the Boltzmann constant, $T$ is the temperature and $p_0$ is the momentum at the roton minimum)\footnote{Experimental measurements of the (first) sound speed in superfluid helium-4 show that $c\sim240$~m~s$^{-1}$, with a weak dependence on the temperature (see, e.g., Ref.~\refcite{vanItterbeek54}). By contrast, we note that $k_\text{B} T_c / p_0 \simeq 14.9$~m~s$^{-1}$ at the critical temperature $T_c\simeq  2.17$~K. Since $c>k_\text{B} T / p_0$, the second condition is thus the most stringent.\cite{khalatnikov1989introduction}} $\rho_{\,\text{n}}$ and $s$ are found to depend only on the temperature, see Chapter 2 of Ref.~\refcite{khalatnikov1989introduction}. Therefore, we must assume for consistency that not only the entropy density $s$ but also the temperature remain unperturbed in the region far from the vortex.  With our assumptions, we have $\delta T=0$, $\delta \rho=0$, $\delta \rho_{\, \text{n}}=0$, $\delta \rho_{\, \text{s}}=0$, and  $\delta s=0$. Contrary to previous studies (see, e.g., Ref.~\refcite{thouless2001vortex}), we do not consider that the fluids are incompressible \emph{everywhere} but only in the region far from the vortex. Let us also recall that these assumptions are only made for the solution obtained when keeping only the first term in the Helmholtz decomposition~\eqref{helmholtz}. In particular, let us emphasize that our analysis does not require the full fluctuations of the temperature, entropy density and mass densities (as obtained by summing the two contributions due to the irrotational and solenoidal solutions respectively) to vanish.

The two-fluid hydrodynamic equations can be found, e.g., in Ref.~\refcite{khalatnikov1989introduction}. In the stationary regime, the linearized hydrodynamic equations (in a frame where the vortex is at rest) thus consist in
\begin{romanlist}[(ii)]
\item the mass conservation
\begin{equation}
\pmb{\nabla} \cdot \delta\pmb{v}=0\, ,
\label{continuity_eq}
\end{equation}
(recalling that the mass current is given by $\rho\,  \pmb{v}=\rho_{\, \text{s}} \, \pmb{ v_{\, \text{s}}}+\rho_{\, \text{n}}\,  \pmb{ v_{\, \text{n}}}$),
\item Euler's equation for the superfluid, 
\begin{equation}
\pmb{\nabla} \left(\delta \mu_{\,  \text{s}} + \pmb{\bar{v}_{\, \text{s}}}\cdot \delta\pmb{v_{\, \text{s}}}  \right) = \pmb{0} \, ,
\label{euler_sup}
\end{equation}
$\mu_{\, \text{s}}$ standing for the superfluid chemical potential per unit mass,  
\item the entropy conservation, 
\begin{equation}
\pmb{\nabla} \cdot \,\delta\pmb{v_{\,\text{n}}} = 0\, ,
\label{cons_entropy}
\end{equation}
\item the total momentum balance equation,
\begin{equation}\label{momentum_balance}
    \nabla_j\delta \Pi^{ij}=0 \, ,
\end{equation}
where, in view of Eqs.~\eqref{irrot_normal_vel} and \eqref{cons_entropy}, we only need consider the non-dissipative part of the momentum-flux tensor given by 
\begin{equation}
\label{stress-energy-tensor}
    \Pi^{ij}=P\delta^{ij} + \rho_{\, \text{s}} v_{\, \text{s}}^i v_{\, \text{s}}^j + \rho_{\, \text{n}}\, v_{\, \text{n}}^i v_{\, \text{n}}^j \, ,
\end{equation}
$P$ being the pressure and we have introduced the Kronecker symbol $\delta^{ij}$, 
\item  the thermodynamic relation, 
\begin{equation}
\delta P = \bar{s}\, \delta T + \bar{\rho}\,  \delta\mu_{\, \text{s}} + \bar{\rho}_{\, \text{n}} \left(\pmb{\bar{v}_{\, \text{n}}} - \pmb{\bar{v}_{\, \text{s}}} \right)   \cdot\delta \left( \pmb{v_{\, \text{n}}} - \pmb{v_{\, \text{s}}} \right)\, .
\label{thermo}
\end{equation}
\end{romanlist}
Let us stress that all the equations presented in this section pertain \emph{only} to the regions far from the vortex and only to the irrotational part of the flows. We make neither assumption on the solenoidal part of the flows nor on the complete flows in the inner regions apart from the global symmetries assumed in Section~\ref{sec:definitions}. 

\subsection{Superfluid versus classical hydrodynamics}
\label{sec:classical-vs-super}
	
The two-fluid hydrodynamic equations cannot be generally reduced to two independent sets of equations for the superflow and for the normal-fluid flow. This stems from the fact that the normal fluid and the superfluid are mutually coupled by nondissipative effects through the introduction of normal and superfluid densities and the thermodynamic relation~\eqref{thermo}. As remarked by Feynman, ``$\rho_{\, \text{n} }$ is a derived concept and is not the density of anything''.\cite{feynman1998statistical} More precisely, $\rho_{\, \text{n} }$ should be viewed as a response function relating the normal momentum per unit entropy (denoted by $\pmb{A}$ in Refs.~\refcite{putterman74superfluid,Geurst1997Iordanskii}), to the normal and superfluid velocities, as follows: 
\begin{equation}
\label{normal_mom}
\pmb{\Theta}=\dfrac{\rho_{\, \text{n} }}{s} (\pmb{v_{\, \text{n}}}-\pmb{v_{\, \text{s}}})\, .
\end{equation}
The physical consequences of Eq.~\eqref{normal_mom} are well summarized by Varoquaux: ``When the superfluid is set into motion, the condensate enforces long-range order and drags the excited states along through the short-range correlations; there is entrainment of the atoms in the fluid by the condensate''.\cite{varoquaux2015} 

The existence of nondissipative couplings between the two fluids appears more clearly in the approach developed by Carter and collaborators, see, e.g., Refs.~\refcite{carter1992equivalence,carter1994canonically,prix2004variational}.  Their consequences in the present context have not been fully appreciated in previous studies, and have been one of the main sources of confusions and speculations.\cite{wexler1998scattering,thouless1999quantized,stone2000Iordanskii} Because the flows of the normal fluid and of the superfluid are intimately related, the contributions of both fluids to the forces acting on a vortex should be treated consistently. This can be achieved starting from the general definition~\eqref{jouko-def}, as shown in the next section. 

\subsection{Hydrodynamic forces on a vortex: explicit formula}
\label{sec:Joukowski-theorem}

Calculating the force from Eq.~\eqref{jouko-def2} happens to be quite intractable in the traditional formulation of Landau and Khalatnikov of the two-fluid hydrodynamic equations summarized in Section~\ref{sec:hydro-eqs}. Instead, we follow 
the approach of Ref.~\refcite{carter2005covariant} based on a fully 4D covariant variational principle that was originally developed for the modeling of the dynamics of neutron stars, the compact remnants of core-collapse supernova explosions.\cite{carter1989} This formulation makes use of geometric concepts such as Killing vectors (see, e.g., Ref.~\refcite{chamel2015}) that have been extremely fruitful in the general theory of relativity (see, e.g., Ref.~\refcite{wald1984}), but that have not been exploited so far in this condensed-matter physics context. The hydrodynamic forces acting on the vortex are found to be expressible as
\begin{equation}
\label{jouko_landau2}
\pmb{\mathcal{F}} =  - \bar{\rho}\  \kappa \, \pmb{\hat{z}} \times \pmb{ \bar{v}} -  \bar{s}\   \mathcal{C}_{\, \text{n}} \, \pmb{\hat{z}} \times \pmb{ \bar{v}_{\, \text{n}}}  + \bar{\rho}_{\, \text{n}}\,  D_{\, \text{n}}\, (\pmb{\bar{v}_{\, \text{n}}} - \pmb{\bar{v}_{\, \text{s}}})\, ,
\end{equation}
where
\begin{equation}
\mathcal{C}_{\, \text{n}} = \oint_\mathcal{C} \pmb{\Theta} \cdot \pmb{\text{d}\ell}\, , 
\label{circulation}
\end{equation}
and 
\begin{equation}
D_{\text{n}} = - \oint_\mathcal{C}  \, \left( \pmb{\hat{z}}\times  \pmb{ v_{\, \text{n}}} \right) \cdot \pmb{\text{d}\ell} \, .
\label{dn-def}
\end{equation}
Full details are given in Appendix~A. Let us recall that Eq.~\eqref{jouko_landau2} accounts only for the force corresponding to the hydrodynamic solution considering only the irrotational part of the normal flow in the Helmholtz decomposition~\eqref{helmholtz}. 

Using Eqs.~\eqref{normal_mom} and \eqref{eq:super_circ}, the normal-fluid momentum circulation integral reduces to 
\begin{equation}
\label{C_n_bis}
\mathcal{C}_{\, \text{n}} = \dfrac{\bar{\rho}_{\, \text{n}}}{\bar{s}} \left( \Gamma_{\, \text{n}} - \kappa \right)\, ,
\end{equation}
where $ \Gamma_{\, \text{n}}$ is the normal-fluid velocity circulation
\begin{equation}\label{gamman-def}
\Gamma_{\, \text{n} } = \oint_\mathcal{C} \pmb{v_{\, \text{n} }} \, \cdot \, \pmb{\text{d}\ell}\, .  
\end{equation}
In view of Eq.~\eqref{C_n_bis}, the normal-fluid (velocity and momentum) circulations  cannot vanish simultaneously, a result that could have hardly been anticipated from analogy with classical hydrodynamics. More surprisingly, Eq.~\eqref{C_n_bis} shows that the normal-fluid circulations are not independent from the circulation quantum $\kappa$, contrary to what could be naively expected. Using Eq.~\eqref{C_n_bis}, the first two terms of Eq.~\eqref{jouko_landau2} can thus be expressed as the sum of the superfluid Magnus force \eqref{superfluid-magnus} and a normal Magnus force 
\begin{equation}\label{normal-Magnus}
\pmb{\mathcal{F}_{\text{nM} }} = - \bar{\rho}_{\, \text{n} }\,\Gamma_{\, \text{n} } \, \pmb{\hat{z}} \times \pmb{\bar{v}_{\, \text{n} }}\, .
\end{equation}
Equations~\eqref{superfluid-magnus} and \eqref{normal-Magnus} were derived in Ref.~\refcite{putterman74superfluid} for the case of a \emph{solid cylinder}  immersed in a superfluid. The authors of Ref.~\refcite{thouless1996tranverse} obtained a similar result from a quantum-mechanical approach. 
Collecting terms and using the identity 
\begin{equation}
   \pmb{V} = - \pmb{\hat{z}} \times (\pmb{\hat{z}} \times \pmb{V}) + (\pmb{V}\cdot \pmb{\hat{z}}) \pmb{\hat{z}}
\end{equation} 
for any vector $\pmb{V}$, the force exerted by the normal fluid on the vortex is thus given by
\begin{equation}\label{eq:Fn}
    \pmb{\mathcal{F}_{\text{n} }}=- \bar{\rho}_{\, \text{n} }\,\Gamma_{\, \text{n} } \, \pmb{\hat{z}} \times \pmb{\bar{v}_{\, \text{n} }} -\bar{\rho}_{\, \text{n} }\, D_{\text{n}}\, \pmb{\hat{z}} \times \bigl[\pmb{\hat{z}} \times ( \pmb{\bar{v}_{\, \text{n} }}-\pmb{\bar{v}_{\, \text{s} }})\bigr] + \bar{\rho}_{\, \text{n} }\, D_{\text{n}} \, \big[( \pmb{\bar{v}_{\, \text{n} }}-\pmb{\bar{v}_{\, \text{s} }})\cdot \pmb{\hat{z}}\bigr] \pmb{\hat{z}}\, .
\end{equation}
Although an axial force of the kind expressed by the last term is generally ignored, some experimental evidence of its existence has been discussed in the literature.\cite{mathieu1984} Moreover, such a force is also expected from microscopic calculations (see, e.g., chapter 8 of Ref.~\refcite{sonin2016}). However, we would like to remind that Eq.~\eqref{eq:Fn} does not represent the complete expression of the force exerted by the normal fluid on the vortex since we only considered the irrotational part of the normal flow in the Helmholtz decomposition~\eqref{helmholtz}.  Therefore, Eq.~\eqref{eq:Fn} cannot be directly compared to experiments at this point.

It can be immediately seen that the normal forces vanish at zero temperature, as expected from the disappearance of the normal fluid. In this limit, the two-fluid hydrodynamic equations reduce to those of classical hydrodynamics for the irrotational flow of an ideal fluid with velocity circulation $\Gamma=\kappa$. Equation~\eqref{jouko_landau2} coincides with the well-known formula of Kutta\cite{kutta1902} and Joukowski\cite{jouko1906}
\begin{equation}
\label{kutta-joukowski}
\pmb{\mathcal{F}_\text{KJ}} =  - \bar{\rho}\  \Gamma \, \pmb{\hat{z}} \times \pmb{ \bar{v}} \, .
\end{equation}
The circulation $\Gamma$ is induced in an ordinary fluid by a solid body, while it is given by the circulation quantum $\kappa$ in the superfluid case. Let us remark that in our derivation we only assumed invariance along the $z$ axis - the section of the body can thus have any shape such as that of an airfoil. As shown in Appendix~\ref{app:KT_singlefluid} following the same approach, Eq.~\eqref{kutta-joukowski} remains valid for an ordinary viscous fluid at finite temperatures. The additional  terms  proportional to $D_{\text{n}}$ in the  superfluid  case  arise from  the existence of nondissipative couplings between the two fluids. Those terms must be retained for the calculations of the complete expression of the force including both the irrotational and solenoidal contributions.

\subsection{Physical relevance of the separate forces}
\label{sec:relevance}

The separate forces in the decomposition \eqref{eq:Fn} are only physically meaningful if the integration contour is chosen in a region far from the vortex, where  Eq.~\eqref{cons_entropy} holds. Indeed, considering two different contours $\mathcal{C}_1$ and $\mathcal{C}_2$, and making use of Stokes' theorem yield 
\begin{equation}\label{Dn-contour}
D_{\, \text{n}}(\mathcal{C}_2) = D_{\, \text{n}}(\mathcal{C}_1)-  \iint_{\mathcal{S}(\mathcal{C}_2)\, \backslash\, \mathcal{S}(\mathcal{C}_1)} \pmb{\nabla} \cdot \pmb{v_{\, \text{n}}}  \, \text{d}S = D_{\, \text{n}}(\mathcal{C}_1)\, ,
\end{equation}
where the integral is over the surface delimited by the two contours as illustrated in Fig.~\ref{fig:surface}. Similarly, the condition~\eqref{irrot_normal_vel} ensures that the normal-fluid velocity circulation $\Gamma_{\, \text{n}}$ is well-defined since 
\begin{eqnarray}
\label{Gamman-contour}
\Gamma_{\, \text{n}}(\mathcal{C}_2)  & = &\Gamma_{\, \text{n}}(\mathcal{C}_1) +  \iint_{\mathcal{S}(\mathcal{C}_2) \, \backslash\, \mathcal{S}(\mathcal{C}_1)} \pmb{\nabla}\times \pmb{v_{\, \text{n}}} \cdot \pmb{\text{d}S} = \Gamma_{\, \text{n}}(\mathcal{C}_1)\, .
\end{eqnarray}


\section{Discussions and conclusions}
\label{sec:conclusions}

We have presented a new approach, originally developed in the astrophysical context  of  compact  stars, to  determine  the  forces acting on a quantized vortex in a superfluid in the framework of Landau-Khalatnikov two-fluid model. 
Considering small perturbations of the asymptotically uniform flows, we need only consider the linearized hydrodynamic equations. The complete solution far from the vortex can thus be obtained by considering separately both the irrotational and solenoidal parts of the normal flow in the Helmholtz decomposition~\eqref{helmholtz}. The general force can then be calculated by summing the two contributions. The additivity of the different force terms is one of the main advantages of our approach compared to previous studies focused on the inner region where the hydrodynamic equations are nonlinear and even break down near and inside the vortex core. As a first step, we focused on the irrotational part of the flows. In this way, we have generalized the Kutta-Joukowski theorem\cite{kutta1902,jouko1906} from classical hydrodynamics and shown that the hydrodynamic force exerted by the superfluid and normal fluid can be expressed in terms of different well-defined contributions including Magnus-type forces, see Eq.~\eqref{jouko_landau2}. Calculations of the complete expression of the force is left for future work. 

The same approach can be also followed to predict the hydrodynamic forces acting on a vortex in systems described by more complicated hydrodynamic equations, such as the A phase of superfluid helium-3 (see, e.g., Ref.~\refcite{vollhardt1990}), starting from Eq.~\eqref{jouko-def} and using the appropriate expression for the momentum-flux tensor. This approach could be also of interest for studying superfluid mixtures of the kind recently produced.\cite{yao2016observation,Matthews1999vortices,schweikhard2004vortex,ferrier2014mixture,roy2017two} Depending on the interactions between the components, different vortex states have been suggested (see, e.g., Refs.~\refcite{mueller2002two,kasamatsu2005vortices,kuopanportti2012exotic} and references therein). If the vortices from the different condensates are arranged into distinct lattices, the force acting on a vortex of component $X$ at zero temperature is easily shown to be given by 
\begin{equation}
\pmb{\mathcal{F}^{\, _X}} =  - \bar{\rho}_{_X}\,   \kappa_{_X} \, \pmb{\hat{z}} \times \pmb{ \bar{v}_{\, _X}}\, ,
\end{equation}
where $\kappa_a$ denotes the corresponding quantum of circulation and we have ignored interactions between vortices. If on the other hand, the vortices overlap, the force will be given by 
\begin{equation}
\label{jouko_12_ii}
\pmb{\mathcal{F}} =  - \sum_{_X} \, \bar{\rho}_{_X}\,   \kappa_{_X} \, \pmb{\hat{z}} \times \pmb{ \bar{v}_{\, _X}}\, . 
\end{equation}
These formulas remain valid if the superfluids are mutually coupled by Andreev-Bashkin effects.\cite{andreev1975three} 

Other possible applications include superconductors and more exotic systems such as neutron stars, whose core is thought to contain a strongly coupled mixture of a neutron superfluid and a type-II proton superconductor.\cite{chamel2017} The conditions prevailing in these astrophysical environments are so extreme that they cannot be reproduced in the laboratory. Studies of the superfluid dynamics in neutron stars have so far relied on bold extrapolations from superfluid helium (see, e.g., Ref.~\refcite{peralta2006}). Existing theories developed for laboratory superfluids, such as that of Ref.~\refcite{kopnin1995spectral}, are not directly applicable to neutron stars since they contain fitting parameters that remain unknown in this context. On the contrary, our approach provides a general theoretical framework for a more realistic description of the vortex dynamics in these compact stars. 

\section*{Acknowledgements}

The authors thank M. Gusakov and A. Sedrakian for valuable discussions. This work was supported by the Fonds de la Recherche Scientifique (Belgium) under grants no. CDR J.0115.18, PDR T.004320  and no. 1.B.410.18F.

\appendix{Derivation of the generalized Kutta-Joukowski formula~\eqref{jouko_landau2}}
\label{app:details-Joukowski}

Following the analysis of Ref.~\refcite{carter2005covariant} within a fully 4D covariant framework, we start by rewriting the momentum-flux tensor in the region far from the vortex in the more convenient form
\begin{equation}
\label{stress-energy-rewritten}
    \Pi^i_j = P\delta^i_j +  n^i p_j  + s^i \Theta_j\, ,
\end{equation}
where $p_j = m v_{\, \text{s}\, j} $ denotes the superfluid momentum per particle and $\Theta_j$~\eqref{normal_mom} the normal momentum per unit entropy, $n^i = n v^i$ is the atomic current ($n = \rho / m$) and $s^i = s  v_{\, \text{n}}^i$ is the entropy current\footnote{The currents are the fundamental objects from which the hydrodynamic equations are derived in the variational formalism developed by Carter and collaborators (see, e.g., Refs.~\refcite{prix2004variational,carter2005covariant,carter1989}). When applied to the two-fluid model, Carter's approach was shown to be fully equivalent to the more traditional Landau-Khalatnikov formulation\cite{carter1992equivalence,carter1994canonically,prix2004variational}, at least in the nondissipative limit considered in Section~\ref{sec:hydro-eqs}.}. Let us recall that we consider here only the irrotational part of the normal velocity in the Helmholtz decomposition and since the normal flow is also incompressible, viscous effects can be ignored. Within our assumptions, $s$ and $\rho_{\, \text{n} }$ are uniform therefore the irrotationality condition~\eqref{irrot_normal_vel} can be equivalently expressed as 
\begin{equation}
\pmb{\nabla}\times \delta \pmb{\Theta}  = \pmb{0}\, . 
\label{irrot_normal}
\end{equation}
The first-order perturbation in the momentum-flux tensor reads 
\begin{equation}
\label{delta_stress_rewritten}
    \delta    \Pi^i_j = \delta   P\, \delta^i_j +  \delta   n^i \, \bar{p}_j + \bar{n}^i\, \delta p_j  + \delta   s^i \, \bar{\Theta}_j + \bar{s}^i \, \delta \Theta_j  \, ,
\end{equation}
where asymptotic uniform values are denoted with a bar. Likewise, the thermodynamic relation~\eqref{thermo} can be rewritten as 
\begin{equation}
\label{delta_thermo_rewritten}
    \delta P = - \bar{n}\,  \delta p_0 - \bar{s}\,  \delta \Theta_0 - \bar{n}^i \, \delta p_i - \bar{s}^i \, \delta \Theta_i\, , 
\end{equation}
where we have introduced the shorthand notations $p_0 = - m\mu_{\,  \text{s}} - m v_{\, \text{s}}^2/2$ and $\Theta_0 = - T - \pmb{\Theta}\cdot\pmb{v_{\, \text{n}}}$, which  in the 4D covariant formulation of Ref.~\refcite{carter2005covariant} correspond to the time-components of the superfluid and normal 4-momenta respectively.

Substituting Eq.~\eqref{delta_thermo_rewritten} into Eq.~\eqref{delta_stress_rewritten}, the hydrodynamic force~\eqref{jouko-def2} experienced by the vortex can thus be decomposed as follows: 
\begin{equation}
\label{jouko-def3}
\mathcal{F}_{ i} = \mathcal{F}^a_{ i} + \mathcal{F}^b_{ i}+ \mathcal{F}^c_{ i} + \mathcal{F}^d_{ i}\, , 
\end{equation}
where 
\begin{equation}
\label{jouko-FA}
\mathcal{F}^a_{ i} = - \oint_\mathcal{C} \left(\bar{n}\,  \delta p_0 +  \bar{s}\,  \delta \Theta_0 \right)\nu_i  \, \text{d}\ell\, , 
\end{equation}
\begin{equation}
\label{jouko-FB}
\mathcal{F}^b_{ i} = \oint_\mathcal{C} \left(  \delta   n^j \, \bar{p}_i  + \delta   s^j \, \bar{\Theta}_i \right)\nu_j  \, \text{d}\ell\, , 
\end{equation}
\begin{equation}
\label{jouko-FC}
\mathcal{F}^c_{i} = \oint_\mathcal{C} \left(   \bar{n}^j\, \delta p_i  - \delta^j_i \bar{n}^k \, \delta p_k\right)\nu_j  \, \text{d}\ell\, , 
\end{equation}
and 
\begin{equation}
\label{jouko-FD}
\mathcal{F}^d_{i} = \oint_\mathcal{C} \left(   \bar{s}^j \, \delta \Theta_i   - \delta^j_i  \bar{s}^k \, \delta \Theta_k \right)\nu_j  \, \text{d}\ell\, .
\end{equation}
These four force terms are evaluated in the following. 

First, let us focus on $\mathcal{F}^a_{i} $. The Euler equation for the superfluid~\eqref{euler_sup} implies that $\nabla_i \, \delta p_0 = 0$, which means that $p_0$ is uniform, i.e., $p_0 = \bar{p}_0$. This evidently leads to $\delta p_0= 0$. Given the assumptions adopted here, the same result actually applies to the normal fluid, i.e., $\delta \Theta_0= 0$. To see this, let us note that 
\begin{align}
    \delta \Theta_0 &= -\delta T - \delta\left(\dfrac{\rho_{\, \text{n} }}{s}\right) (\bar{v}_{\, \text{n}}^i-\bar{v}_{\, \text{s}}^i)\, \bar{v}_{\, \text{n}\, i} \nonumber \\
    &- \dfrac{\bar{\rho}_{\, \text{n} }}{\bar{s}} \left[ (\delta v_{\, \text{n}}^i-\delta v_{\, \text{s}}^i)\, \bar{v}_{\, \text{n}\, i} +  ( \bar{v}_{\, \text{n}}^i-\bar{v}_{\, \text{s}}^i)\, \delta v_{\, \text{n}\, i} \right]\, ,
\end{align}
 which follows from the very definition of $\Theta_0$. The first two terms appearing on the right-hand side vanish as a consequence of our assumptions that $\delta T = \delta s = \delta \rho_{\, \text{n}}=0$ (see Section~\ref{sec:hydro-eqs}). Besides, the linearized total momentum balance equation~\eqref{momentum_balance} ensures that the remaining terms (inside brackets) are also zero. Therefore, one simply has $\mathcal{F}^a_{i} = 0$. The vanishing of $\delta p_0$ and $\delta\Theta_0$ arises naturally in the 4D covariant approach, as a consequence of the irrotationality conditions~\eqref{irrot_sup} and \eqref{irrot_normal}, together with the invariance of the fluids under the action of the Killing vector associated with the time translation symmetry.\cite{carter2005covariant}
 
Introducing the coefficients $\mathcal{D}$ and  $\mathcal{D}_{\,\text{n}}$ as
\begin{equation}
    \mathcal{D} = \oint_\mathcal{C} n^j \, \nu_j  \, \text{d}\ell \ \ \text{and} \ \  \mathcal{D}_{\,\text{n}} = \oint_\mathcal{C} s^j \, \nu_j  \, \text{d}\ell\, ,
\end{equation}
the force term $\mathcal{F}_{ i}^b$ can be rewritten as 
\begin{equation}
\mathcal{F}_{ i}^b = \bar{p}_i \, \mathcal{D}  + \bar{\Theta}_i \,  \mathcal{D}_{\,\text{n}}\, ,
\end{equation}
where we have used the fact that $\delta \mathcal{D} = \mathcal{D} - \bar{\mathcal{D}} = \mathcal{D} $ and  $\delta \mathcal{D}_{\,\text{n}} = \mathcal{D}_{\,\text{n}} - \bar{\mathcal{D}}_{\,\text{n}}=\mathcal{D}_{\,\text{n}}$, since the uniform background values necessarily vanish. Using Stokes' theorem, $\mathcal{D}$ and $\mathcal{D}_{\,\text{n}}$ can be equivalently expressed as 
\begin{equation}
\mathcal{D}=-\iint_{\mathcal{S}(\mathcal{C})} \pmb{\nabla} \cdot \left(n\,  \pmb{v}\right)   \text{d}S\, , 
\end{equation}
and 
\begin{equation}\label{eq:def-mathcalDn}
\mathcal{D}_{\, \text{n}}=-\iint_{\mathcal{S}(\mathcal{C})} \pmb{\nabla} \cdot \left(s\,  \pmb{v_{\, \text{n}}}\right)   \text{d}S\, , 
\end{equation}
where the integrals are over the surface $\mathcal{S}\left(\mathcal{C}\right)$  delimited by the contour $\mathcal{C}$ and d$S$ is the corresponding surface element. In view of~\eqref{continuity_eq} and~\eqref{cons_entropy}, these two integrals  are independent of the chosen contour in the region of interest, such that the quantities $\mathcal{D}$ and $\mathcal{D}_{\,\text{n}}$ are well-defined. Furthermore, since the continuity equation~\eqref{continuity_eq} must hold \textit{everywhere} (i.e., not only in the region far from the vortex core, but also in the innermost vortex regions), one has $\mathcal{D} =0$. The force term $\pmb{\mathcal{F}^b}$ therefore reduces to
\begin{equation}\label{eq:Fb}
\pmb{\mathcal{F}^b} = \pmb{\bar{\Theta}} \, \mathcal{D}_{\, \text{n}}=\bar{\rho}_{\, \text{n}}\, D_{\, \text{n}}\, (\pmb{\bar{v}_{\, \text{n}}} - \pmb{\bar{v}_{\, \text{s}}})  \, ,
\end{equation}
where $ D_{\, \text{n}}= \mathcal{D}_{\, \text{n}}/\bar{s}$. 

As shown in Ref.~\refcite{carter2005covariant}, the existence of another Killing vector associated with the longitudinal invariance along the vortex line and the irrotationality conditions~\eqref{irrot_sup} and \eqref{irrot_normal} lead to $\hat{z}^j \delta p_j = 0$ and $\hat{z}^j \delta \Theta_j = 0$. This result can be alternatively derived by writing explicitly the irrotationality conditions~\eqref{irrot_sup} and \eqref{irrot_normal} in cylindrical coordinates and considering that the momenta are independent of $z$. Defining $\perp^i_j$ as the operator of projection orthogonal to the vortex line, i.e., $\perp^i_j = \delta^i_j - \hat{z}^i\hat{z}_j$, one thus has 
\begin{equation}
\perp^i_j \delta p^j = \delta p^i, \hskip0.5cm \perp^i_j \delta \Theta^j = \delta \Theta^i\, .
\end{equation}
The force term $\mathcal{F}^c_i$  can therefore be recast as 
\begin{equation}
    \mathcal{F}^c_{ i}   = \oint_\mathcal{C} \bar{n}^j\delta p_k \left[ \perp^k_i   \,  \nu_j -\perp^k_j\,  \nu_i \right] \, \text{d}\ell\, .
\end{equation}
Introducing the unit vector $\pmb{\beta}$ along the contour such that $\nu_i = -\varepsilon_{ijk}\beta^j\hat{z}^k$ (see Fig.~\ref{fig:surface}), where $\varepsilon_{ijk}$ is the Levi-Civita symbol, and making use of the identity
\begin{equation}
    \perp^k_i\varepsilon_{jlm} - \perp^k_j \varepsilon_{ilm} = \perp^k_l \varepsilon_{jim}\, , 
\end{equation}
the third term in \eqref{jouko-def3} can be equivalently written as
\begin{equation}
    \mathcal{F}^c_{ i}   = - \varepsilon_{ijm} \hat{z}^j \bar{n}^m  \delta \mathcal{C}_{\, \text{s}}\, ,
\end{equation}
where the superfluid momentum integral $\mathcal{C}_{\, \text{s}} $ is given by 
\begin{equation}
\mathcal{C}_{\, \text{s}} =   \oint_\mathcal{C}\pmb{p}\cdot \pmb{\text{d}\ell}\, ,  
\end{equation}
with $\pmb{\text{d}\ell} = \text{d}\ell\, \pmb{\beta}$.
Since the background value $\bar{\mathcal{C}}_{\, \text{s}} $ vanishes and using the fact that the superfluid momentum circulation integral is quantized, i.e., one has
\begin{equation}
\label{eq:super_circ}
   \mathcal{C}_{\, \text{s}} = m \, \kappa
\end{equation}
 in the presence of a single vortex line,
the force term $\pmb{\mathcal{F}^c}$ finally reads
\begin{equation}
  \pmb{\mathcal{F}^c}   = - \bar{\rho}\,  \kappa\,  \pmb{\hat{z}}\times \pmb{\bar{v}}\, .
\end{equation}
Similar arguments also apply to the normal fluid, with the only but important difference that the normal momentum circulation integral 
\begin{equation}
   \mathcal{C}_{\, \text{n}} = \oint_\mathcal{C}\pmb{\Theta}\cdot \pmb{\text{d}\ell}\, ,  
\end{equation}
is not quantized. The use of the Stokes' theorem, i.e.,
\begin{equation}
\mathcal{C}_{\, \text{n}} =\iint_{\mathcal{S}(\mathcal{C})} \pmb{\nabla}\times \pmb{\Theta} \cdot \pmb{\text{d}S}\, , 
\end{equation}
in conjunction with the irrotationality condition~\eqref{irrot_normal} ensure that $\mathcal{C}_{\, \text{n}}$ is well-defined. The last force term $\pmb{\mathcal{F}^d}$ thus reads
\begin{equation}
  \pmb{\mathcal{F}^d}   = - \bar{s}\,   \mathcal{C}_{\, \text{n}} \,  \pmb{\hat{z}}\times \pmb{\bar{v}_{\, \text{n}}}\, .
\end{equation}

Collecting terms in Eq.~\eqref{jouko-def3} leads to the expression given in Eq.~\eqref{jouko_landau2} for the Kutta-Joukowski force experienced by the vortex line.

\appendix{Derivation of the Kutta-Joukowski formula for the force felt by a solid body immersed in an ordinary fluid}
\label{app:KT_singlefluid}

To check the validity of our approach, we rederive here the classical expression of the Kutta-Joukowski formula~\eqref{kutta-joukowski} for the force felt by a longitudinally-invariant solid body plunged into an ordinary fluid, using a procedure similar to that followed in Appendix~A for the treatment of a superfluid at finite temperatures. 

As in the superfluid case, we focus on the irrotational part of the flow far from the solid body  (using a Helmholtz decomposition similar to that given in Eq.~\eqref{helmholtz}), such that we have 
\begin{equation}
\pmb{\nabla}\times \delta \pmb{v}  = \pmb{0}\, . 
\label{irrot_singlefluid}
\end{equation}
We also neglect the first-order fluctuations of the mass density $\rho$ and entropy density~$s$.
These assumptions also imply that $\delta T = 0$ since the entropy density of an incompressible fluid only depends on the temperature. The linearized continuity equation with $\delta \rho =0$ thus leads to $\pmb{\nabla}\cdot \delta \pmb{v}=0$. Viscous effects can thus be ignored in the region far from the body considered here (but note that we do not neglect viscous effects in the vicinity of the body). The momentum-flux tensor reduces to~(see, e.g., Ref.~\refcite{prix2004variational}) 
\begin{equation}
\label{stress-energy-singlefluid}
    \Pi^i_j = P\delta^i_j +  n^i p_j \, ,
\end{equation}
where $n^i = n v^i$ is the atomic current ($n=\rho/m$ being the atom number density, $m$ the atomic mass, and $v^i$  the fluid velocity) and  $p_j = m v_{\, j} $ is the fluid momentum per particle. By contrast with the superfluid case, the entropy does not contribute to the momentum-flux tensor although there exists an associated current $s^i$. The reason lies in the vanishing of the momentum per unit entropy $\Theta_i$, which is nonzero in the superfluid case only due to nondissipative couplings between the two fluids (see Secs.~IV.A and IV.C of Ref.~\refcite{prix2004variational}). 

The first-order perturbation in the momentum-flux tensor reads 
\begin{equation}
\label{delta_stress_singlefluid}
    \delta    \Pi^i_j = \delta   P\, \delta^i_j +  \delta   n^i \, \bar{p}_j + \bar{n}^i\, \delta p_j  \, ,
\end{equation}
where asymptotic uniform values are denoted with a bar. The thermodynamic relation governing the system, i.e., 
\begin{equation}
    \label{thermo_singlefluid}
   \delta P =   \bar{s}\, \delta T + \bar{\rho} \, \delta \mu\, ,
\end{equation}
where $\mu$ denotes the fluid chemical potential per unit mass, can be  rewritten as 
\begin{equation}
\label{delta_thermo_rewritten_singlefluid}
    \delta P = - \bar{n}\,  \delta p_0 - \bar{s}\,  \delta \Theta_0 - \bar{n}^i \, \delta p_i\, ,
\end{equation}
where we have introduced the shorthand notations $p_0 = - m \mu - m v^2/2$ and $\Theta_0 = - T$, which in the 4D covariant formulation of Ref.~\refcite{carter2005covariant} correspond to the time-components of the fluid and entropy 4-momenta, respectively.  

Substituting Eq.~\eqref{delta_thermo_rewritten_singlefluid} into Eq.~\eqref{delta_stress_singlefluid}, the hydrodynamic force~\eqref{jouko-def2} experienced by the solid body can thus be decomposed as follows: 
\begin{equation}
\label{jouko-def3_singlefluid}
\mathcal{F}_{ i} = \mathcal{F}^a_{ i} + \mathcal{F}^b_{ i}+ \mathcal{F}^c_{ i} \, , 
\end{equation}
where 
\begin{equation}
\label{jouko-FA_singlefluid}
\mathcal{F}^a_{ i} = - \oint_\mathcal{C} \left(\bar{n}\,  \delta p_0 +  \bar{s}\,  \delta \Theta_0 \right)\nu_i  \, \text{d}\ell\, , 
\end{equation}
\begin{equation}
\label{jouko-FB_singlefluid}
\mathcal{F}^b_{ i} = \oint_\mathcal{C} \,  \delta   n^j \, \bar{p}_i \, \nu_j  \, \text{d}\ell\, , 
\end{equation}
\begin{equation}
\label{jouko-FC_singlefluid}
\mathcal{F}^c_{i} = \oint_\mathcal{C} \left(   \bar{n}^j\, \delta p_i  - \delta^j_i \bar{n}^k \, \delta p_k\right)\nu_j  \, \text{d}\ell\, . 
\end{equation}

Let us first concentrate on  $\mathcal{F}^a_{i} $. The Euler equation governing the fluid motion (we recall here that viscous effects do not play any role far from the body given the assumptions followed here) implies that $\nabla_i \, \delta p_0 = 0$, which means that $p_0$ is uniform, i.e., $p_0 = \bar{p}_0$. This evidently leads to $\delta p_0= 0$. Given the assumptions adopted here, we also have $\delta \Theta_0= -\delta T = 0$. Therefore, one simply has $\mathcal{F}^a_{i} = 0$, as in the superfluid case.

Introducing 
\begin{equation}
    \mathcal{D} = \oint_\mathcal{C} n^j \, \nu_j  \, \text{d}\ell \, ,
\end{equation}
the force term $\mathcal{F}_{ i}^b$ can be rewritten as 
\begin{equation}
\mathcal{F}_{ i}^b = \bar{p}_i \, \mathcal{D} \, ,
\end{equation}
where we have used the fact that $\delta \mathcal{D} = \mathcal{D} - \bar{\mathcal{D}} = \mathcal{D} $. Using Stokes' theorem, $\mathcal{D}$ can be equivalently expressed as 
\begin{equation}
\mathcal{D}=-\iint_{\mathcal{S}(\mathcal{C})} \pmb{\nabla} \cdot \left(n\,  \pmb{v}\right)   \text{d}S\, , 
\end{equation}
where the integral is over the surface $\mathcal{S}\left(\mathcal{C}\right)$ delimited by the contour $\mathcal{C}$ and d$S$ is the corresponding surface element. Since the continuity equation $\pmb{\nabla} \cdot \left(n\,  \pmb{v}\right) = 0$ must hold \textit{everywhere} (i.e., not only in the region far from the solid body, but also in the innermost regions), one has $\mathcal{D} =0$. The force term $\pmb{\mathcal{F}^b}$ therefore vanishes.

The existence of another Killing vector associated with the longitudinal invariance along the solid body and the irrotationality condition~\eqref{irrot_singlefluid} lead to $\hat{z}^j \delta p_j = 0$. Using a similar procedure as in Appendix~A, the force term $\mathcal{F}^c_i$  can therefore be recast as 
\begin{equation}
  \pmb{\mathcal{F}^c}   = - \bar{\rho}\,  \Gamma \,  \pmb{\hat{z}}\times \pmb{\bar{v}}\, ,
\end{equation}
where the coefficient $ \Gamma$ is given by the velocity circulation
\begin{equation}
\Gamma =   \oint_\mathcal{C}\pmb{v}\cdot \pmb{\text{d}\ell}\, .  
\end{equation}
Note that the irrotationality condition~\eqref{irrot_singlefluid} ensures that $ \Gamma$ is well-defined.

Collecting terms in Eq.~\eqref{jouko-def3_singlefluid} leads to the well-known expression~\eqref{kutta-joukowski} for the Kutta-Joukowski force experienced by a solid body immersed in an ordinary fluid. The presence of additional terms in the superfluid case arises from the existence of nondissipative couplings between the two fluids.

\section*{References}
 

\begin{thebibliography}{10}

\bibitem{essmann1967}
U.~{Essmann} and H.~{Tr{\"a}uble}, {\em Phys. Lett. A} {\bf 24}, 526
  (1967).

\bibitem{yarmchuk1979}
E.~J. {Yarmchuk}, M.~J.~V. {Gordon} and R.~E. {Packard}, {\em Phys. Rev. Lett.} {\bf 43}, 214 (1979).

\bibitem{abo-shaeer2001}
J.~R. {Abo-Shaeer}, C.~{Raman}, J.~M. {Vogels} and W.~{Ketterle}, {\em Science}
  {\bf 292}, 476 (2001).

\bibitem{zwierlein2005}
M.~W. {Zwierlein}, J.~R. {Abo-Shaeer}, A.~{Schirotzek}, C.~H. {Schunck} and
  W.~{Ketterle}, {\em Nature} {\bf 435}, 1047 (2005).

\bibitem{yao2016observation}
X.-C. {Yao}, H.-Z. {Chen}, Y.-P. {Wu}, X.-P. {Liu}, X.-Q. {Wang}, X.~{Jiang},
  Y.~{Deng}, Y.-A. {Chen} and J.-W. {Pan}, {\em Phys. Rev. Lett. } {\bf
  117}, p. 145301 (2016).

\bibitem{donnelly2005}
R.~J. {Donnelly}, {\em Quantized vortices in helium II} (Cambridge University
  Press, 2005).

\bibitem{mangin2017}
P.~{Mangin} and R.~{Kahn}, {\em {Superconductivity}} (Springer, 2017).

\bibitem{khalatnikov1989introduction}
I.~M. Khalatnikov, {\em An introduction to the theory of superfluidity}
  (Perseus Books, 1989).

\bibitem{wexler1997}
C.~{Wexler}, {\em Phys. Rev. Lett. } {\bf 79}, 1321 (1997).

\bibitem{vinen1961}
W.~F. {Vinen}, {\em Proc. R. Soc. Lond. Ser. A} {\bf
  260}, 218 (1961).

\bibitem{whitmore1968}
S.~C. {Whitmore} and W.~{Zimmermann}, {\em Phys. Rev.} {\bf 166}, 181
  (1968).

\bibitem{zieve1993}
R.~J. {Zieve}, J.~D. {Close}, J.~C. {Davis} and R.~E. {Packard}, {\em J. Low. Temp. Phys.} {\bf 90}, 243 (1993).

\bibitem{putterman74superfluid}
S.~J. Putterman, {\em Superfluid hydrodynamics} (Amsterdam, North-Holland
  Publishing Co.; New York, American Elsevier Publishing Co.,
  Inc.(North-Holland Series in Low Temperature Physics. Volume 3), 1974).

\bibitem{thouless1996tranverse}
D.~J. {Thouless}, P.~{Ao} and Q.~{Niu}, {\em Phys. Rev. Lett. } {\bf 76},
  3758 (1996).

\bibitem{kopnin2002vortex}
N.~B. {Kopnin}, {\em Rep. Prog. Phys.} {\bf 65}, 1633 (2002).

\bibitem{sonin2010}
E.~B. {Sonin}, {\em J. Phys. A Math. Gen.} {\bf 43}, p.
  354003 (2010).

\bibitem{barenghi1983friction}
C.~F. {Barenghi}, R.~J. {Donnelly} and W.~F. {Vinen}, {\em J. Low. Temp. Phys.} {\bf 52}, 189 (1983).

\bibitem{bevan1995vortex}
T.~D.~C. {Bevan}, A.~J. {Manninen}, J.~B. {Cook}, A.~J. {Armstrong}, J.~R.
  {Hook} and H.~E. {Hall}, {\em Phys. Rev. Lett. } {\bf 74}, 750 (1995).

\bibitem{bevan1997vortex}
T.~D.~C. {Bevan}, A.~J. {Manninen}, J.~B. {Cook}, H.~{Alles}, J.~R. {Hook} and
  H.~E. {Hall}, {\em J. Low. Temp. Phys.} {\bf 109}, 423 (1997).

\bibitem{mathieu1984}
P.~{Mathieu}, B.~{Pla{\c c}ais} and Y.~{Simon}, {\em Phys. Rev. B} {\bf
  29}, 2489 (1984).

\bibitem{sonin2016}
E.~B. {Sonin}, {\em {Dynamics of Quantised Vortices in Superfluids}} (Cambridge
  University Press, 2016).

\bibitem{mathieu1980hydro}
P.~{Mathieu} and Y.~{Simon}, {\em Phys. Rev. Lett. } {\bf 45}, 1428
  (1980).

\bibitem{thouless2001vortex}
D.~J. {Thouless}, M.~R. {Geller}, W.~F. {Vinen}, J.-Y. {Fortin} and S.~W.
  {Rhee}, {\em Phys. Rev. B} {\bf 63}, p. 224504 (2001).

\bibitem{sonin2001LNP}
E.~B. Sonin, Magnus {Force}, {Aharonov}-{Bohm} {Effect}, and {Berry} {Phase} in
  {Superfluids}, in {\em Quantized Vortex Dynamics and Superfluid
  Turbulence\/},  eds. C.~F. Barenghi, R.~J. Donnelly and W.~F. Vinen, Lecture
  Notes in Physics, Vol.~571 (Springer-Verlag, Berlin, 2001), pp. 131-137. 

\bibitem{kutta1902}
W.~M. {Kutta}, {\em Illustr. A\"eron. Mitt. } {\bf 6}, 133
  (1902).

\bibitem{jouko1906}
N.~E. {Joukowski}, {\em Bull. Inst. A\'erodyn. Koutchino}
  {\bf 1}, 51 (1906).
  
\bibitem{aris1989}
R. {Aris}, {\em Tensors, Vectors, and the Basic Equations of Fluid Mechanics} (Dover Publications, New York, 1989).   

\bibitem{elwell1967}
D.~L. {Elwell} and H.~{Meyer}, {\em Phys. Rev. } {\bf 164}, 245 (1967).

\bibitem{harrislowe68}
R.~F. {Harris-Lowe} and K.~A. {Smee}, {\em Phys. Lett. A} {\bf 28}, 246
  (1968).

\bibitem{meijdenberg1961}
C.~J.~N. {Van den Meijdenberg}, K.~W. {Taconis} and R.~{De Bruyn Ouboter}, {\em
  Physica} {\bf 27}, 197 (1961).

\bibitem{vanItterbeek54}
A.~{Van Itterbeek} and G.~{Forrez}, {\em Physica} {\bf 20}, 133 (1954).

\bibitem{Geurst1997Iordanskii}
J.~A. {Geurst} and H.~{van Beelen}, {\em Physica A Stat. Mech. Appl.} {\bf 237}, 1 (1997).

\bibitem{feynman1998statistical}
R.~{Feynman}, {\em Statistical mechanics: a set of lectures (advanced book
  classics)} (Westview Press, 1998).

\bibitem{varoquaux2015}
E.~{Varoquaux}, {\em Rev. Mod. Phys. } {\bf 87}, 803 (2015).

\bibitem{carter1992equivalence}
B.~{Carter} and I.~M. {Khalatnikov}, {\em Phys. Rev. D} {\bf 45}, 4536
  (1992).

\bibitem{carter1994canonically}
B.~{Carter} and I.~M. {Khalatnikov}, {\em Rev. Math. Phys.} {\bf
  6}, 277 (1994).

\bibitem{prix2004variational}
R.~{Prix}, {\em Phys. Rev. D} {\bf 69}, p. 043001 (2004).

\bibitem{wexler1998scattering}
C.~{Wexler} and D.~J. {Thouless}, {\em Phys. Rev. B} {\bf 58}, R8897
  (1998).

\bibitem{thouless1999quantized}
D.~J. {Thouless}, P.~{Ao}, Q.~{Niu}, M.~R. {Geller} and C.~{Wexler}, {\em
  Int. J. Mod. Phys.  B} {\bf 13}, 675 (1999).

\bibitem{stone2000Iordanskii}
M.~{Stone}, {\em Phys. Rev. B} {\bf 61}, 11780 (2000).

\bibitem{carter2005covariant}
B.~{Carter} and N.~{Chamel}, {\em Int. J. Mod. Phys.  D}
  {\bf 14}, 717 (2005).

\bibitem{carter1989}
B.~Carter, {Covariant} {Theory} of {Conductivity} in {Ideal} {Fluid} or {Solid}
  {Media}, in {\em {Relativistic Fluid Dynamics}\/},  eds. A.~M. Anile and
  Y.~Choquet-Bruhat, Lecture Notes in Mathematics, Vol.~1385 (Springer-Verlag,
  Berlin, 1989), pp. 1-64. 

\bibitem{chamel2015}
N.~{Chamel}, {\em Int. J. Mod. Phys.  D} {\bf 24}, 1550018
  (2015).

\bibitem{wald1984}
R.~M. {Wald}, {\em {General relativity}} (University of Chicago Press, 1984).

\bibitem{vollhardt1990}
D.~{Vollhardt} and P.~{Wolfle}, {\em {The Superfluid Phases of Helium 3}}
  (Taylor and Francis, 1990).

\bibitem{Matthews1999vortices}
M.~R. Matthews, B.~P. Anderson, P.~C. Haljan, D.~S. Hall, C.~E. Wieman and
  E.~A. Cornell, {\em Phys. Rev. Lett. } {\bf 83}, 2498 (1999).

\bibitem{schweikhard2004vortex}
V.~{Schweikhard}, I.~{Coddington}, P.~{Engels}, S.~{Tung} and E.~{Cornell},
  {\em Phys. Rev. Lett. } {\bf 93}, p. 210403 (2004).

\bibitem{ferrier2014mixture}
I.~{Ferrier-Barbut}, M.~{Delehaye}, S.~{Laurent}, A.~T. {Grier}, M.~{Pierce},
  B.~S. {Rem}, F.~{Chevy} and C.~{Salomon}, {\em Science} {\bf 345}, 1035
  (2014).

\bibitem{roy2017two}
R.~{Roy}, A.~{Green}, R.~{Bowler} and S.~{Gupta}, {\em Phys. Rev. Lett. }
  {\bf 118}, p. 055301 (2017).

\bibitem{mueller2002two}
E.~J. {Mueller} and T.-L. {Ho}, {\em Phys. Rev. Lett. } {\bf 88}, p.
  180403 (2002).

\bibitem{kasamatsu2005vortices}
K.~{Kasamatsu}, M.~{Tsubota} and M.~{Ueda}, {\em Int. J. Mod. Phys. B} {\bf 19}, 1835 (2005).

\bibitem{kuopanportti2012exotic}
P.~{Kuopanportti}, J.~A.~M. {Huhtam{\"a}ki} and M.~{M{\"o}tt{\"o}nen}, {\em
  Phys. Rev. A} {\bf 85}, p. 043613 (2012).

\bibitem{andreev1975three}
A.~F. Andreev and E.~P. Bashkin, {\em Soviet J. Exp. Theor. Phys.} {\bf 42}, p. 164 (1975).

\bibitem{chamel2017}
N.~{Chamel}, {\em J. Astrophys. Astron.} {\bf 38}, p.~43
  (2017).

\bibitem{peralta2006}
C.~{Peralta}, A.~{Melatos}, M.~{Giacobello} and A.~{Ooi}, {\em Astrophys. 
  J.} {\bf 651}, 1079 (2006).

\bibitem{kopnin1995spectral}
N.~B. {Kopnin}, G.~E. {Volovik} and {\"U}.~{Parts}, {\em EPL (Europhys. Lett.)} {\bf 32}, 651 (1995).

\end{thebibliography}

\end{document}